\documentclass[runningheads]{llncs}
\usepackage[T1]{fontenc}
\usepackage{graphicx}
\usepackage{subcaption}
\usepackage{amssymb,amsmath}
\usepackage[ruled,linesnumbered]{algorithm2e}
\usepackage{breqn}
\begin{document}
\title{Minsum Problem for Discrete and Weighted Set Flow on Dynamic Path Network\thanks{This research was partly conducted during Bubai Manna's and Bodhayan Roy's visit to The University of Tokyo. The visit was hosted by Prof. Reiji Suda and was supported by the JST Sakura Science Program. Vorapong Suppakitpaisarn was partially supported by KAKENHI Grant 23H04377. The authors would like to thank the reviewers for their comments, which significantly improved this paper.}}
\author{Bubai Manna \inst{1} \and Bodhayan Roy \inst{1} \and Vorapong Suppakitpaisarn\inst{2}}
\institute{IIT Kharagpur, Kharagupur, India \and The University of Tokyo, Tokyo, Japan}
\titlerunning{Minsum Problem for Discrete Flow on Dynamic Path Network}

\maketitle              
\begin{abstract}\vspace{-0.5cm}
In this research, we examine the minsum flow problem in dynamic path networks where flows are represented as discrete and weighted sets. The minsum flow problem has been widely studied for its relevance in finding evacuation routes during emergencies such as earthquakes. However, previous approaches often assume that individuals are separable and identical, which does not adequately account for the fact that some groups of people, such as families, need to move together and that some groups may be more important than others. To address these limitations, we modify the minsum flow problem to support flows represented as discrete and weighted sets. We also propose a 2-approximation pseudo-polynomial time algorithm to solve this modified problem for path networks with uniform capacity.
\end{abstract}\vspace{-0.5cm}
\section{Introduction}
\label{sec:typesetting-summary}

Flow problems on dynamic graphs \cite{ford1958constructing} are considered by many researchers (e.g. \cite{higashikawa2022almost,klinz2004minimum}) because of many reasons. One of the reasons is their relevance in finding evacuation routes during emergencies such as earthquakes or fires \cite{higashikawa2019survey}. In those applications, we aim to move persons in the ways that they arrive at aiding facilities as soon as possible.

A common objective function for those problems is minmax, which aims to minimize the time until all persons arrive at facilities. In this work, however, we consider another common objective function called minsum, which aims to minimize the summation of time that each individual needs for their trips.

\begin{figure}
     \centering
     \begin{subfigure}[b]{0.35\textwidth}
         \centering
         \includegraphics[width=\textwidth]{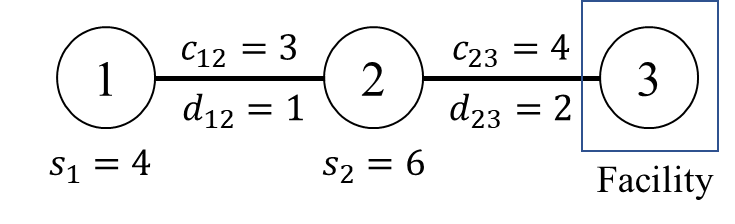}
         \caption{}
     \end{subfigure}
     \begin{subfigure}[b]{0.35\textwidth}
         \centering
         \includegraphics[width=\textwidth]{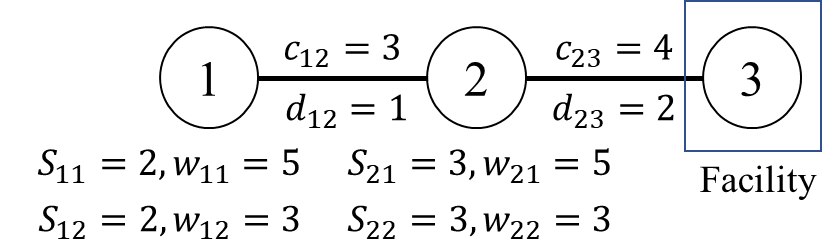}
         \caption{}
     \end{subfigure}
        \caption{(a) An instance of the dynamic flow problem (b) An instance of the problem for discrete and weighted set flow on dynamic network}
\label{fig1}
\end{figure}

\begin{example}In Figure \ref{fig1}a, there are 4 people at node 1 and 6 people at node 2. These 10 people need to be transported to the aid facility at node 3. Both edges have capacity constraints: a maximum of 3 people can be moved on the edge between nodes 1 and 2 in one unit of time, and a maximum of 4 people can be moved on the edge between nodes 2 and 3 in one unit of time. It takes 1 unit of time to travel from node 1 to node 2 and 2 units of time to travel from node 2 to node 3. At time 1, we can move 3 people from node 1 to node 2 and 4 people from node 2 to node 3. This leaves 1 person at node 1, 5 people at node 2, and 4 people in the middle of the edge between nodes 2 and 3. At time 2, the remaining person at node 1 is moved to node 2, and 4 people at node 2 are moved to node 3. This results in 4 people from node 2 arriving at the facility within 2 units of time, 4 people arriving within 3 units of time, and 2 people arriving within 4 units of time. The maximum time was 4 units of time. The summation of times was $4 \times 2 + 4 \times 3 + 2 \times 4 = 28$. The move which we discussed here minimized both the maximum time and the summation.
\end{example}

It has been shown that both objective functions of flow problems can be solved using time-expanded networks \cite{ford1958constructing,fu2016clustering}. However, these temporal graphs can be exponentially large in relation to the input size, making the algorithm pseudo-polynomial. 
For minmax problems, polynomial-time algorithms have been developed for paths \cite{bhattacharya2017improved} and trees \cite{treeminmax}. There are also FPTAS for general graphs when the number of facilities is constant \cite{belmonte2015polynomial}. In contrast, minsum problems have only been shown to have polynomial-time algorithms for path graphs \cite{benkoczi2020minsum}. 

All known algorithms assume that individuals are distinct and identical, meaning that we can move any number of people over a particular edge as long as the total number does not exceed the edge's capacity. However, this may not always be possible in practice. For example, some groups of people, such as families, must be moved together, and some groups may require emergency aid and should therefore be given higher priority. These considerations must be taken into account when determining how to move people from one location to another.

\subsection{Our Contributions}

In short, we modify the minsum flow problem to support flows represented as discrete and weighted sets. We also propose a 2-approximation pseudo-polynomial time algorithm to solve this modified problem for path networks with uniform capacity. 

We illustrate the ideas of the modified problem in the following example.
\begin{example}
In Figure 1b, there are two groups of people at node 1, each with 2 people. The weight of the first group is 5, while the weight of the second group is 3. There are also two groups of 3 people at node 2, with weights of 5 and 3, respectively. We refer to the group with size $S_{ij}$ and weight $w_{ij}$ as $G_{ij}$. At time 1, we move group $G_{11}$ from node 1 to node 2 and group $G_{21}$ from node 2 to node 3. Before time 2, group $G_{12}$ is at node 1, groups $G_{11}$ and $G_{22}$ are at node 2, and group $G_{21}$ is in the middle of the edge between nodes 2 and 3. At time 2, we move group $G_{12}$ to node 2 and group $G_{11}$ to node 3. At time 3, we move group $G_{12}$ from node 2 to node 3, and at time 4, we move group $G_{22}$. As a result, group $G_{21}$ arrives at time 2, group $G_{11}$ arrives at time 3, group $G_{12}$ arrives at time 4, and group $G_{22}$ arrives at time 5. The weighted summation of arrival time is then $2w_{21} + 3w_{11} + 4w_{12} + 5w_{22} = 52$.
\end{example}

It is clear that the modified problem is harder than the original version. Indeed, we can show that it is NP-hard by a reduction to the partition problem. The formal definition of this problem with its NP-hardness proof can be found in Section 2.

We discuss in Section 3 that when we have two nodes, our problem is equivalent to the weighted minsum bin packing problem \cite{epstein2007minimum,epstein2018min}. To support the case that we have more than two nodes, we need to derive a bin-packing algorithm that can support items with different arrival times. Suppose that $t_i$ is the arrival time of item $i$. The item cannot be packed in the first $(t_i - 1)$-th bag. 
We show that the algorithm is a 2-approximation. 

As there is PTAS proposed in \cite{epstein2018min} for the minsum bin packing problem, one may think that we can extend that PTAS to support items with arrival times. Unfortunately, by the requirement that we cannot insert particular items in some bags, we strongly believe that the extension is not straightforward. We are aiming to give that extension as our future work. 

In Section 4, we extend the bin packing algorithm presented in Section 3 to address our main problem. We demonstrate that the extended algorithm is a 2-approximation when all capacities are uniform, and there is only one facility. It is worth noting that several works in dynamic network flows also make this assumption of uniform capacities \cite{higashikawa2014minimax,kamiyama2006efficient} and a single facility \cite{fujie2021minmax,higashikawa2015minimax}.

\section{Problem Definitions}
In this section, we define our problem called minsum problem for discrete and weighted set flow on a dynamic path network (MS-DWSF). 

Consider a path graph with $n$ nodes, denoted by $P_n = (V = \{1,\dots,n\}, E = \{\{i,i+1\}: 1 \leq i \leq n - 1 \})$. Each node $i$ has $m_i$ sets of persons to evacuate. Those sets of persons are denoted by $G_{i,1}, \dots, G_{i,m_i}$. For group $G$, the size of $G$ is denoted by $S(G) \in \mathbb{Z}_{+}$ and the weight of $G$ is denoted by $w(G) \in \mathbb{Z}_{+}$. The capacity of all edges is $C \in \mathbb{Z}_{+}$. Each edge $e$ has distance $d(e) \in \mathbb{Z}_{+}$, which is the time that persons need to move between two terminals of the edge. 

Suppose that the single aiding facility is located at $a \in V$. People originally at node $i < a$ must move in a direction that increases the node number they are at, while people originally at node $i > a$ must move in a direction that decreases the node number they are at in any optimal solution.

Let us denote the collection of groups that are at node $i$ at time $t$ by $\mathcal{S}_i^{(t)}$. We select from $\mathcal{S}_i^{(t)}$ which groups to be sent along the edge $\{i, i + 1\}$ for $i < a$ and along the edge $\{i - 1, i\}$ for $i > a$. We denote the collection of groups that we choose to send by $D_i^{(t)}$. The summation of group sizes in $D_i^{(t)}$ must not be larger than $C$, i.e. $\sum\limits_{G \in D_i^{(t)}} S(G) \leq C$.

For $t = 0$, we have $\mathcal{S}_i^{(0)} = \{G_{i,1}, \dots, G_{i,m_i}\}$ for all $i$.
Let denote $A_i^{(t)}$ be a collection of groups arriving at $i$ from node $i - 1$ at time $t$ and denote $B_i^{(t)}$ be a collection of groups arriving at $i$ from node $i + 1$ at time $t$. We have $A_i^{(t)} = D_{i-1}^{(t - d(\{i - 1, i\}))}$ when $1 < i \leq a \text{ and } t \geq d(\{i-1,i\})$ and $A_i^{(t)} = \emptyset$ otherwise. Similarly, $B_i^{(t)} = D_{i+1}^{(t - d(\{i, i + 1\}))}$ when $a \leq i < n \text{ and } t \geq d(\{i,i + 1\})$ and $B_i^{(t)} = \emptyset$ otherwise.
Then, $\mathcal{S}_i^{(t)} = \mathcal{S}_i^{(t - 1)} \backslash D_i^{(t)} \cup A_i^{(t)} \cup B_i^{(t)}$.

The arrival time of $G$, denoted by $\alpha(G)$ is the earliest time that the group is at $a$, i.e. $\min \{t: G \in \mathcal{S}_a^{(t)}\}$. In the MS-DWSF, we aim to minimize $\sum\limits_G w(G)\alpha(G)$. We show that the problem is NP-hard in Appendix.

\section{Minsum Bin Packing Problem for Weighted Items with Different Ready Times}

To address the MS-DWSF problem, we first introduce a related problem called the minsum bin packing problem for weight items with different ready times (MS-BPWRT). In this section, we present a 2-approximation pseudo-polynomial time algorithm for the MS-BPWRT problem. We will then use the solution obtained from this algorithm to develop a 2-approximation pseudo-polynomial time algorithm for the MS-DWSF in the following section.

\subsection{Definition of MS-BPWRT}
The MS-BPWRT problem can be defined in the following definition:
\begin{definition}[MS-BPWRT($\tau$)] Given a collection of groups $\mathcal{G} = \{G_1, \dots, G_m\}$. Each group $G_i$ has size $S(G_i)$, weight $w(G_i)$, and ready time $\tau(G_i)$. We find a way to pack those groups into a set of bins $B_1, \dots, B_T  \subseteq \mathcal{G}$ with capacity $C$ with the following constraints:
    1) $\bigcup\limits_{1 \leq j \leq m} B_j = \{G_1, \dots, G_m\},$ 
    2) $B_j \cap B_{j'} = \emptyset$ for $j \neq j'$, 
    3) For all $1 \leq j \leq T$, $\sum\limits_{G \in B_j} S(G) \leq C$, and
    4) Denote $t(G) = j$ when $G \in B_j$, we must have $t(G) \geq \tau(G)$.
We aim to minimize $\sum\limits_G w(G) t(G)$.
\end{definition}
When $w(G) = 1$ for all $G$ and we do not have the fourth constraint, the MS-BPWRT is equivalent to the minsum bin packing problem \cite{epstein2018min}. We use some ideas from the minsum bin packing problem to provide an algorithm and prove the approximation ratio for the MS-BPWRT.

We have included the weight of group $G$, denoted as $w(G)$, in the problem formulation because we recognize that different groups may have varying levels of importance. The ready time, $\tau(G)$, signifies that group $G$ cannot be placed in any bin with an index less than $\tau(G)$. In other words, group $G$ is not ready to be inserted until time $\tau(G)$.

\subsection{Approximation Algorithm for MS-BPWRT}

The approximation algorithm for MS-BPWRT is described in Algorithm \ref{alg:01}. The collection $\mathcal{G}$ contains groups that have not been placed in any bin, while the collection $\mathcal{G}_j'$ is a candidate set for bin $B_j$. If any remaining group can be considered a candidate for $B_j$ by replacing the code in Line 3 with $\mathcal{G}_j' \leftarrow \mathcal{G}'$, then Algorithm \ref{alg:01} becomes the next fit decreasing algorithm \cite{rhee1987probabilistic}, based on the ratio $w(G)/S(G)$. It is worth noting that the minsum bin packing algorithm in \cite{epstein2018min} uses the next fit increasing algorithm based on $s(G)$ (or the next fit decreasing algorithm based on $1/s(G)$). The criteria for the next fit algorithm is how we apply the weights $w(G)$ to the minsum bin packing problem.

\begin{algorithm} 
\small
\caption{2-Approximation Algorithm for MS-BPWRT}\label{alg:01}
\KwIn{  Collection of all groups $\mathcal{G}$, size of each group $S: \mathcal{G} \rightarrow \mathbb{Z}_{+}$, weight of each group $w: \mathcal{G} \rightarrow \mathbb{Z}_{+}$, ready time of each group $\tau: \mathcal{G} \rightarrow \mathbb{Z}_{+}$, capacity of each bin $C \in \mathbb{Z}_{+}$}
\KwOut{ Groups in bins $B_1, \dots, B_T \subseteq \mathcal{G}$}
 $\mathcal{G'} \leftarrow \mathcal{G}$; $j \leftarrow 1$\\
\While{$\mathcal{G'} \neq \emptyset$}{
    $\mathcal{G}'_j \leftarrow \{G \in \mathcal{G'}: \lceil (\tau(G) - 1) / 2 \rceil \times 2 + 1 \leq j \}$\\
    $G' \leftarrow \arg \max\limits_{G \in \mathcal{G}'_j} w(G) / S(G)$\\
    \eIf{$\sum\limits_{G \in B_j} S(G) + S(G') > C$}
    {
        $j \leftarrow j + 1$\\
    }
    {
        $B_j \leftarrow B_j \cup \{G'\}, \mathcal{G'} \leftarrow \mathcal{G'} \backslash \{G'\}$
    }
}
\end{algorithm}

We consider the ready times at Line 3 of the algorithm. The collection $\mathcal{G}_j'$ is the collection of groups that have not been added to any bin of which the ready time $\tau(G)$ satisfies $\lceil (\tau(G) - 1) / 2 \rceil \times 2 + 1 \leq j$. We know that $\lceil (\tau(G) - 1) / 2 \rceil \times 2 + 1 = \tau(G)$ when $\tau(G)$ is odd and $\lceil (\tau(G) - 1) / 2 \rceil \times 2 + 1 = \tau(G) + 1$ when $\tau(G)$ is even. Recall that $\mathcal{G}_j'$ is the candidate to be added to $B_j$. For $G$ such that $\tau(G)$ is odd, we add $G$ to the candidate set of $B_j$ for any $j \geq \tau(G)$ that matches with the ready time constraint. On the other hand, for $G$ such that $\tau(G)$ is even, we do not add $G$ to the candidate set of $B_j$ for $j = \tau(G)$, but add only when $j \geq \tau(G) + 1$. Informally, we delay the addition of $G$ by one bin here.

\subsection{Proof for Approximation Ratio}

We prove that the algorithm in the previous subsection is a two-approximation algorithm for the MS-BPWRT problem. First, we define the relaxed version of MS-BPWRT in the following definition:
\begin{definition}[MS-BPWRT-REAL($\tau$)] Suppose we have a collection of groups $\mathcal{G} = {G_1, \dots, G_m}$, where each group $G_i$ has a size of $S(G_i)$, a weight of $w(G_i)$, and a ready time of $\tau(G_i)$. We are given a capacity $C$. For each $1 \leq i \leq m$ and $1 \leq j \leq T$, we find $x_{ij} \in [0,1]$ such that 1) $\sum\limits_j x_{ij} = 1$ for all $i$, 2) $\sum\limits_i S(G_i) x_{ij} \leq C$ for all $j$, and, 3) for each $x_{ij} > 0$, we must have $j \geq \tau(G_i)$.
Our goal is to minimize $\sum\limits_{i,j} j \cdot w(G_i) \cdot x_{ij}$.
\end{definition}

Informally speaking, in the MS-BPWRT-REAL problem, each group can be partially assigned to each bin. The variable $x_{ij}$ represents the proportion of group $G_i$ assigned to bin $B_j$. Let $OPT(\tau), OPT_R(\tau)$ be the optimal value of the MS-BPWRT($\tau$) and MS-BPWRT-REAL($\tau$) problems. We have the following properties:
\begin{proposition}
$OPT_R(\tau) \leq OPT(\tau)$. \label{prop6}
\end{proposition}
\begin{proof}
    Let $B_1^*, \dots, B_T^*$ be an optimal solution of MS-BPWRT($\tau$), $T^*(G_i) = j$ if $G_i \in B_j^*$, and let $x_{ij}' = 1$ if $G_i \in B_j^*$ and $x_{ij}' = 0$ otherwise. We know that $\langle x_{ij}' \rangle_{i,j}$ is a solution of MS-BPWRT-REAL($\tau$) because
        1) $\sum_j x'_{ij} = 1$ for all $i$ because each $G_i$ is a member of exactly one bin by constraints 1) and 2) of MS-BPWRT($\tau$),
        2) For all $j$, $\sum\limits_i S(G_i)x'_{ij} = \sum\limits_{G_i \in B^*_j} S(G_i) \leq C$ by the third constraint of MS-BPWRT($\tau$),
        3) When $x'_{ij} > 0$, $G_i \in B_j^*$ and, by the fourth constraint of MS-BPWRT($\tau$), $j \geq \tau(G)$.
    The objective value of $\langle x_{ij}' \rangle_{i,j}$ is 
        $\sum_{i,j} w(G_i) (j \cdot x_{ij}') = \sum_{i} w(G_i) \sum_j (j \cdot x_{ij}') = \sum_{i} w(G_i) T^*(G_i) = OPT(\tau).$
   We then know that there is a solution of MS-BPWRT-REAL($\tau$) with objective value $OPT(\tau)$. The optimal value of MS-BPWRT-REAL($\tau$) must not be larger than $OPT(\tau)$, i.e. $OPT_R(\tau) \leq OPT(\tau).$ \qed
\end{proof}

\begin{proposition}
    Let $\tau,\tau'$ be a function such that $\tau(G) \leq \tau'(G)$ for all $G \in \mathcal{G}$. Then, $OPT_R(\tau) \leq OPT_R(\tau')$. \label{prop7}
\end{proposition}
\begin{proof}
    Let $\langle x^*_{ij} \rangle_{i,j}$ be an optimal solution of the MS-BPWRT-REAL($\tau'$) problem. It is straightforward to show that $\langle x^*_{ij} \rangle_{i,j}$ satisfies the first and the second constraints of MS-BPWRT-REAL($\tau$). Also, because for each $x_{ij} > 0$, we have $j \geq \tau'(G_i) \geq \tau(G_i)$, we know that $\langle x^*_{ij} \rangle_{i,j}$ also satisfies the third constraint of MS-BPWRT-REAL($\tau$), and is a feasible solution of MS-BPWRT-REAL($\tau$). 
    
    The objective value of $\langle x^*_{ij} \rangle$ is $OPT_R(\tau')$. We then know that there is a solution of MS-BPWRT-REAL($\tau$) with objective value $OPT_R(\tau')$. The optimal value of MS-BPWRT-REAL($\tau$) must not be larger than $OPT_R(\tau')$, i.e. $OPT_R(\tau) \leq OPT_R(\tau').$    \qed
\end{proof}

Denote a solution from Algorithm \ref{alg:01} by $B_1', \dots, B_T'$. Let $x_{ij}' = 1$ when $G_i \in B'_{2j - 1} \cup B'_{2j}$. It is clear that $\langle x_{ij}' \rangle_{i,j}$ is not a feasible solution of MS-BPWRT-REAL($\tau'$). We prove a property of $\langle x_{ij}' \rangle_{i,j}$ in the following proposition:
\begin{proposition}
    \label{prop8}
    For all $j$ such that $B'_{2j} \neq \emptyset$, $\sum_i S(G_i) x_{ij}' > C$.
\end{proposition}
\begin{proof}
At Line 3 of Algorithm \ref{alg:01}, we define the set $\mathcal{G}_j'$. We can observe that both $\mathcal{G}_{2j - 1} = \{G \in \mathcal{G}': \tau(G) \leq 2j - 1\}$ and $\mathcal{G}_{2j} = \{G \in \mathcal{G}': \tau(G) \leq 2j - 1\}$ are defined in the same manner. This implies that, for any integer $k$, even if we increase $k$ from $2k - 1$ to $2j$ at Line 6, the set of groups considered remains unchanged.  

Let $G'$ be the first element added to the bin $B_{2j}'$. It is a group in $G'_{2j} \backslash (B'_1 \cup \dots \cup B'_{2j - 1})$ which maximizes $w(G)/S(G)$. 

Since the set of groups considered for bins $B'_{2j-1}$ and $B'_{2j}$ are the same, $G'$ must have already been considered for inclusion in $B'_{2j - 1}$. However, it was not added to $B'_{2j - 1}$ because doing so would result in $\sum\limits_{G \in B'_{2j - 1}} S(G) + S(G') > C$.

Therefore, by the definition of $x_{ij}'$, we have
$\sum_i S(G_i) \cdot x_{ij}' = \sum_{G \in B'_{2j - 1}} S(G) + \sum_{G\in B'_{2j}} S(G) \geq \sum_{G \in B'_{2j - 1}} S(G) + S(G') > C.$
This completes the proof. \qed
\end{proof}

Next, we define a problem called weight maximization problem (WM) as follows:
\begin{definition}[WM($\tau',C_1, \dots, C_{T'}$)] Suppose we have a collection of groups $\mathcal{G} = {G_1, \dots, G_m}$, where each group $G_i$ has a size of $S(G_i)$, a weight of $w(G_i)$, and a ready time of $\tau(G_i)$. We are given a capacity $C$. For each $1 \leq i \leq m$ and $1 \leq j \leq T'$, we find $x_{ij} \in [0,1]$ such that:
1) $\sum\limits_j x_{ij} \leq 1$ for all $i$, 
2) $\sum\limits_i S(G_i) x_{ij} \leq C_j$ for all $j$, and,
3) for each $x_{ij} > 0$, we must have $j \geq \tau(G_i)$.
We aim to maximize $\sum\limits_{i,j} w(G_i) \cdot x_{ij}$.
\end{definition}

Let $C'_j = \sum_i S(G_i)x_{ij}'$, and let $\tau'$ be a function such that, for all $G_i \in \mathcal{G}$, $\tau'(G_i) = \lceil \tau(G_i) / 2 \rceil$.
We then can show the following property:
\begin{proposition}
    For all $1 \leq T' \leq T$, $\langle x_{ij}' \rangle_{j \leq T',i}$ is an optimal solution of WM($\tau',C'_1, \dots, C'_{T'}$). \label{prop10}
\end{proposition}
\begin{proof}
    We prove this proposition by induction on $T'$.

Let us examine the scenario where $T' = 1$. Remember from Algorithm \ref{alg:01} that the bins $B_1$ and $B_2$ contain groups from the set $\{G \in \mathcal{G}: \tau(G) = 1\}$ that maximize the ratio $w(G)/S(G)$. Using the greedy algorithm, any collection of groups $\mathcal{D} \subseteq \{G \in \mathcal{G}: \tau'(G) = 1\}$ with $\sum\limits_{G \in \mathcal{D}} S(G) \leq C'_1$ must satisfy $\sum\limits_{G \in \mathcal{D}} w(G) \leq \sum\limits_{G \in B_1 \cup B_2} w(G) = \sum\limits_{i} w(G_i) \cdot x_{i1}'$. Therefore, we can deduce that the sequence $\langle x_{ij}' \rangle_{j=1,i}$ represents an optimal solution for WM($\tau',C'_1$).

   Next, let us assume the proposition holds true for all $T' \leq \mathsf{T}$. We will assume, aiming for a contradiction, that the sequence $\langle x_{ij}' \rangle_{j \leq \mathsf{T}, i}$ does not represent an optimal solution for WM($\tau', C'_1, \dots, C'_{\mathsf{T}}$). Let us say an optimal solution for WM($\tau', C'_1, \dots, C'_{\mathsf{T}}$) is represented by the sequence $\langle x_{ij}^* \rangle_{j \leq \mathsf{T}, i}$.

Based on our assumption that the sequence $\langle x'_{ij} \rangle_{j \leq \mathsf{T} - 1, i}$ is an optimal solution for WM($\tau', C'_1, \dots, C'_{\mathsf{T} - 1}$), it follows that $\sum\limits_{j \leq \mathsf{T} - 1, i} w(G_i) \cdot x'_{ij} \geq \sum\limits_{j \leq \mathsf{T} - 1, i} w(G_i) \cdot x_{ij}^*$. To satisfy the condition $\sum\limits_{j \leq \mathsf{T}, i} w(G_i) \cdot x'_{ij} \geq \sum\limits_{j \leq \mathsf{T}, i} w(G_i) \cdot x_{ij}^*$, it is necessary to have $\sum\limits_i w(G_i) x_{i\mathsf{T}}' < \sum\limits_i w(G_i) x_{i\mathsf{T}}^*$.
    
    Recall from the construction that $x_{ij}' \in \{0,1\}$ for all $i$. To have $\sum\limits_i w(G_i) x_{i\mathsf{T}}' < \sum\limits_i w(G_i) x_{i\mathsf{T}}^*$, there must be $i^*$ such that $x_{{i^*}\mathsf{T}}' = 0$, $x_{{i^*}\mathsf{T}}^* > 0$, and $w(G_{i^*})/S(G_{i^*}) > w(G_{i})/S(G_{i})$ for all $i$ such that $x_{i\mathsf{T}}' = 1$.
    
    Consider the case that $x'_{i^*\mathsf{T}'} = 1$ for some $\mathsf{T}' < \mathsf{T}$.  Then, in the solution $\langle x_{ij}^* \rangle_{j \leq \mathsf{T},i}$, we move $G_{i^*}$ to the bin $B'_{2\mathsf{T}' - 1} \cup B'_{2\mathsf{T}'}$. 
    Let $s = \min \{S(G_{i^*})x^*_{i^*\mathsf{T}}, C_{\mathsf{T'}}' - \sum\limits_i S(G_{i}) x^*_{i\mathsf{T}'}\}$. 
    If $s > 0$, we have more spaces to put the group $G_{i^*}$. We then decrease the value of $x^*_{i^*,\mathsf{T}}$ by $s/S(G_{i^*})$ and increase the value of $x^*_{i^*,\mathsf{T}'}$ by the same value.
    
    If $C'_{\mathsf{T'}} - \sum\limits_i S(G_{i}) x^*_{i\mathsf{T}'} = 0$, there is no space left in the bin $B_{\mathsf{T}'}$. We then have to swap $G_{i^*}$ with some other groups. There is an item $i'$ such that $x^*_{i'\mathsf{T}'} > 0$ while $x'_{i'\mathsf{T}'} = 0$. Let $s = \min\{S(G_{i'})x^*_{i'\mathsf{T}'}, S(G_{i^*})x^*_{i^*\mathsf{T}}\}$. We can update the value of $\langle x^*_{ij} \rangle_{i,j}$ in the following ways:
    1) decrease the value of $x^*_{i^*\mathsf{T}}$ by $s/S(G_{i^*})$,
        2) decrease the value of $x^*_{i'\mathsf{T}'}$ by $s/S(G_{i'})$,
        3) increase the value of $x^*_{i^*\mathsf{T}'}$ by $s/S(G_{i^*})$, and
        4) increase the value of $x^*_{i'\mathsf{T}}$ by $s/S(G_{i'})$.
    Informally, we exchange $s$ units of $G_{i^*}$ in bin $\mathsf{T}$ with an equal mass of $G_{i'}$ in bin $\mathsf{T}'$. This updated result continues to be a feasible solution for WM$(\tau',C'_1, \dots, C'_T)$, and the objective value remains unchanged.

    We can iterate the update in the previous paragraph until there is no $i^*$ such that $x'_{i^*\mathsf{T}'} = 1$ for some $\mathsf{T}'$. It is sufficient to only consider the case when, for all such $i^*$, we have not included the group $G_{i^*}$ to bins $B_1, \dots, B_{2\mathsf{T} - 2}$. However, by the assumption that $w(G_{i^*})/S(G_{i^*}) > w(G_{i})/S(G_{i})$ for all $i$ such that $x_{i\mathsf{T}}' = 1$, the greedy algorithm must have already included the group $i^*$ to the bin $B_{2\mathsf{T} - 1}$. This gives $x'_{i^*\mathsf{T}} = 1$, which contradicts our assumption that $x'_{i^*\mathsf{T}} = 0$. \qed
\end{proof}    

From the next proposition, let us consider the problem WM$(\tau', C_1, \dots, C_{T'})$ where $C_1 = C_2 = \cdots = C_{T'} = C$. We denote the optimal solution of the problem by $OPT_{\text{WM}}(T')$.

\begin{proposition}
    For all $1 \leq T' \leq T$, $OPT_{\text{WM}}(T') \leq \sum\limits_{j \leq T', i} w(G_i) \cdot x'_{ij}$. \label{prop11}
\end{proposition}
\begin{proof}
Let $T'$ be such that $B'_{2T'} \neq \emptyset$. Using the definition of $C_j$ and Proposition \ref{prop8}, we have $C_j = \sum\limits_i S(G_i) x'_{ij} > C$. An optimal solution for WM($\tau', C, \dots, C$) is a feasible solution for WM($\tau', C_1', \dots, C_{T'}'$). Therefore, the objective value of an optimal solution for WM($\tau', C_1', \dots, C_{T'}'$), which is $\sum\limits_{j \leq T', i} w(G_i) \cdot x'_{ij}$ according to Proposition \ref{prop10}, cannot be less than $OPT_{\text{WM}}(T')$.

When $B'_{2T'} = \emptyset$, it implies that all items have been allocated to the bins $B'_1, \dots, B'_{2T'-1}$. In this case, the value of $\sum\limits_{j \leq T',i} w(G_i) x_{ij}'$ is equal to $\sum\limits_{i} w(G_i)$, which is greater than or equal to the sum of the weights of any feasible solution. Hence, we have $\sum\limits_{j \leq T',i} w(G_i) x_{ij}' = OPT_{\text{WM}}(T')$. \qed
\end{proof}

The next lemma gives a relationship between the sequence $\langle x'_{ij} \rangle_{i,j}$ and the MS-BPWRT-REAL problem.
\begin{lemma}
$OPT_R(\tau') \geq \sum\limits_{i,j} j \cdot w(G_i) \cdot x'_{ij}$ \label{lemma12}
\end{lemma}
\begin{proof}
    Let $\mathcal{X}$ be a collection of all feasible solutions of the MS-BPWRT-REAL($\tau'$), and let $W = \sum\limits_{i} w(G_i)$. By Proposition \ref{prop11}, we have that
   $OPT_R(\tau') = \min\limits_{\langle x_{ij} \rangle_{i,j} \in \mathcal{X}} \sum\limits_j \sum\limits_i j \cdot w(G_i) \cdot x_{ij}
    = \min\limits_{\langle x_{ij} \rangle_{i,j} \in \mathcal{X}} \sum\limits_{T'} \sum\limits_{j \geq j', i} w(G_i) \cdot x_{ij}
    \geq \sum\limits_{T'} \min\limits_{\langle x_{ij} \rangle_{i,j} \in \mathcal{X}} \left[ W - \sum_{j < T',i} w(G_i) \cdot x_{ij} \right]
    = \sum\limits_{T'} \left[W - \max\limits_{\langle x_{ij} \rangle_{i,j} \in \mathcal{X}} \sum\limits_{j < T',i} w(G_i) \cdot x_{ij}  \right]
    \geq \sum\limits_{T'} \left[W - \sum_{j < T',i} w(G_i) \cdot x'_{ij}  \right] 
    \\ = \sum\limits_{T'} \sum\limits_{j \geq j',i} w(G_i) \cdot x_{ij}'
    = \sum\limits_{i,j} j \cdot w(G_i) \cdot x'_{ij}$. \qed
\end{proof}

We are now ready to prove the main theorem of this section. 
\begin{theorem}
    The bin $B'_1, \dots, B'_T$ obtained from Algorithm \ref{alg:01} is a 2-approximation solution for MS-BPWRT.
\end{theorem}
\begin{proof}
    Let $SOL$ be an objective value of $B'_1, \dots, B'_T$. We have that
         $SOL = \sum_{j} \sum_{G_i \in B'_j} j \cdot w(G_i) \leq \sum_{j} \sum_{G_i \in B'_{2j - 1} \cup B'_{2j}} 2j \cdot w(G_i)
        \leq 2 \cdot \sum_{i,j} j \cdot w(G_i) \cdot x'_{ij} \leq 2 \cdot OPT_R(\tau') \leq 2 \cdot OPT_R(\tau) \leq 2 \cdot OPT(\tau).$
    The inequality at Line 3 of the chain is obtained from the definition of $\langle x_{ij}' \rangle_{i,j}$. The inequality at Line 4 is obtained from Lemma \ref{lemma12}, the inequality at Line 5 is obtained from Proposition~\ref{prop7}, and the inequality at Line 6 is obtained from Proposition~\ref{prop6}. \qed
\end{proof}

\section{Approximation Algorithm for MS-DWSF}

In this section, we will develop an approximation algorithm for our main problem, MS-DWSF, utilizing the findings presented in the previous section.

\subsection{Algorithm}

Our two-approximation algorithm for the MS-DWSF is shown in Algorithm \ref{alg:02}.  The algorithm addresses congestion on the busiest edge. Specifically, when the destination node is denoted as $a$, the most congested edges are $\{a-1,a\}$ and $\{a,a+1\}$. To tackle this issue, we can examine separate strategies for each of these edges. It is worth noting that the concepts behind both strategies are the same, so we will only elaborate on the approach for edge $\{a-1,a\}$ here.

\begin{algorithm}
\small
\caption{2-Approximation Algorithm for MS-DWSF}\label{alg:02}
\KwIn{  1) A path graph $V = \{1,\dots,n\}$ and $E = \{\{i,i+1\}:1 \leq i \leq n -1\}$, \\ 2) for each $1 \leq i \leq n$, a set of groups $S_i^{(0)} = \{G_{i,1}, \dots, G_{i,m_i}\}$, \\ 3) for each group $G_{i,j}$, the size of $G_{i,j}$ denoted by $S(G_{i,j})$ and the weight of $G_{i,j}$ denoted by $w(G_{i,j})$. \\ 4) for each edge $e \in E$, a distance of $e$ denoted by $d(e)$. \\ 5) the capacity of all edges denoted by $C$. \\
6) the destination node denoted by $a \in V$}
\KwOut{ Groups to move from node $i$ at time $t$ denoted by $D_i^{(t)}$}
For $i < a$, let $d(i, a - 1) = \sum\limits_{v = i}^{a - 2} d(\{v,v+1\})$. \\ 
Execute Algorithm \ref{alg:01} under the following conditions: $\mathcal{G}$ is a set of groups $\{G_{i,j}: i < a\}$, where $\tau(G_{i,j}) = d(i, a - 1)$, and the size and weight of each group correspond to the input parameters of MS-DWSF. Suppose that the output of the algorithm is $B_1', \dots, B_T'$ \\
$D_i^{(t)} \leftarrow B'_{t + d(i, a - 1)} \cap \{G_{i',j}: i' \leq i\}$ for all $i < a$ and $t$.\\
For $i > a$, let $d(i, a + 1) = \sum\limits_{v = i}^{a} d(\{v,v+1\})$. \\
Execute Algorithm \ref{alg:01} under the following conditions: $\mathcal{G}$ is a set of groups $\{G_{i,j}: i > a\}$, where $\tau(G_{i,j}) = d(i, a + 1)$, and the size and weight of each group correspond to the input parameters of MS-DWSF. Suppose that the output of the algorithm is $B_1', \dots, B_T'$ \\
$D_i^{(t)} \leftarrow B'_{t + d(i, a + 1)} \cap \{G_{i',j}: i' \geq i\}$ for all $i > a$ and $t$.\\
\end{algorithm}

To transmit all groups in the set $\{G_{i,j}: i < a\}$ through the edge $\{a - 1, a\}$, we rely on the results of Algorithm 1 (denoted by $B'_1, \dots, B'_T$) to determine the appropriate timing for each item. At time $T'$, items within bin $B_{T'}$ are dispatched along the $\{a - 1, a\}$ edge. 

The MS-DWSF constraint requires that all groups $G_{i,j} \in B_{T'}$ be present at node $a-1$ during transmission. To satisfy this condition, if $i \leq a-2$, the group is sent from node $a-2$ to $a-1$ at the time $T' - d(\{a-1,a-2\})$. Similarly, if $i \leq v$, the group is sent from node $v$ to $a-1$ at time $T' - d(v,a-1),$ following the same idea. 

The collection of groups transmitted from node $i$ at time $t$ is obtained by taking the intersection of $B'_{t + d(i,a-1)}$ with $\{G_{i',j}: i' \leq i\}$, as assigned in Line 3 of the algorithm.

Since group $G_{i,j}$ is initially located at node $i$, it cannot reach node $a-1$ before time $d(i,a-1)$. Thus, it is not possible to assign group $G_{i,j}$ to bin $B_j$ for $j < d(i,a-1)$. This is why we set $\tau(G_{i,j}) = d(i,a-1)$ in Line 2 of the algorithm.

\subsection{Feasibility and Approximation Ratio}

In this subsection, we show that Algorithm \ref{alg:02} always gives a feasible solution. Then, we show that it is a two-approximation ratio for MS-DWSF.

\begin{theorem}
    $\langle D_i^{(t)} \rangle_{i,t}$ in Algorithm \ref{alg:02} is a feasible solution to MS-DWSF.
\end{theorem}

\begin{proof}
To show that the solution of Algorithm \ref{alg:02} is feasible, we need to show that, for all $i$ and $t$, $\sum\limits_{G \in D_i^{(t)}} S(G) \leq C$ and $D_i^{(t)} \subseteq S_i^{(t)}$. The first inequality can be shown by the fact that $D_i^{(t)} \subseteq B'_{t + d(i,a-1)}$ and $\sum\limits_{G \in B'_{t + d(i,a-1)}} S(G) \leq C$ by the constraint of MS-BPWRT.

We will now demonstrate that $D_i^{(t)} \subseteq S_i^{(t)}$. Suppose we have a group $G_{i',j} \in D_i^{(t)}$. Since $G_{i',j} \in D_i^{(t)}$, we have $G_{i',j} \in B'_{t + d(i, a - 1)}$.
By the MS-BPWRT constraint, we know that $G_{i',j}$ cannot be assigned to any bin $B_{T'}$ for $T' < t + d(i, a - 1)$. As a result, $G_{i',j}$ is not in $D_i^{(t')}$ for any $t' < t$. Hence, for $i' = i$, we conclude that $G_{i',j} \in S_i^{(t)}$. 

For $i' < i$, we have that $G_{i',j} \in B'_{t + d(i,a-1)} = B'_{t - d(\{i - 1, i\}) + d(i - 1, a - 1)}$. For $t' = t - d(\{i - 1, i\})$, we have $G_{i',j} \in B'_{t' + d(\{i - 1, a - 1\})}$ and $G_{i',j} \in D_{i - 1}^{(t')}$. By the problem definition of MS-DWSF, we know that $G_{i',j} \in S_i^{(t)}$ when $G_{i',j} \in D_{i - 1}^{(t - d(\{i - 1, i\})}$.  \qed
\end{proof}

The next theorem will show that Algorithm \ref{alg:02} is a two-approximation algorithm for MS-DWSF.

\begin{theorem}
    Let $\alpha'(G)$ be the time that $G$ arrives at the node $a$ in Algorithm \ref{alg:02}, and let $OPT_D$ be an optimal solution of the MS-DWSF problem. We have that $\sum_G w(G) \alpha'(G) \leq 2 \cdot OPT_D$. 
\end{theorem}
\begin{proof}
    We can construct a feasible solution of MS-BPWRT from an optimal solution of MS-DWSF by setting $B_{T'}$ to $D_{a - 1}^{(T')}$ for all $T'$. Let $W_D = d(\{a-1,a\}) \sum\limits_G w(G)$. As a group $G_{i,j} \in D_{a - 1}^{(T')}$ arrives at the destination node $a$ at time $T' + d(\{a-1,a\})$, we have that 
    $\sum\limits_{T'} \sum\limits_{G \in B_{T'}} T' \cdot  w(G) = \sum\limits_{T'} \sum\limits_{G \in B_{T'}} (\alpha(G) - d(\{a - 1, a\})) \cdot  w(G) = \sum\limits_G \alpha(G) w(G) - W_D \hiderel{=} OPT_D - W_D.$
    If $OPT_B$ is an optimal value of MS-BPWRT, we have that $OPT_B \leq OPT_D - W_D$.

    Algorithm~\ref{alg:01} gives a solution of MS-BPWRT of which the objective function, denoted by $SOL_B$, is not larger than $2 \cdot OPT_B$. From that solution, we can construct a solution of MS-DWSF using Line 3 of Algorithm \ref{alg:02}. The objective value of the MS-DWSF solution, denoted by $SOL_D$, is $SOL_B + W_D$. We then obtain that
        $SOL_D = SOL_B + W_D \leq 2SOL_B + W_D \leq 2 OPT_D.$ \qed
\end{proof}

\section{Conclusion}
This paper presents an extension of the minsum bin packing problem, which considers items with varying ready times and weights. We propose a 2-approximation algorithm for this new problem and apply it to develop an evacuation method for non-separable groups of individuals.
At present, our algorithm is limited to path graphs with a single destination. However, we are actively working on expanding its capabilities to handle multiple destinations and non-path network structures.

%
%
%
\bibliographystyle{splncs04}
\bibliography{aaim}

\begin{thebibliography}{10}
\providecommand{\url}[1]{\texttt{#1}}
\providecommand{\urlprefix}{URL }
\providecommand{\doi}[1]{https://doi.org/#1}

\bibitem{belmonte2015polynomial}
Belmonte, R., Higashikawa, Y., Katoh, N., Okamoto, Y.: Polynomial-time approximability of the k-sink location problem. arXiv preprint arXiv:1503.02835  (2015)

\bibitem{benkoczi2020minsum}
Benkoczi, R., Bhattacharya, B., Higashikawa, Y., Kameda, T., Katoh, N.: Minsum k-sink problem on path networks. Theoretical Computer Science  \textbf{806},  388--401 (2020)

\bibitem{bhattacharya2017improved}
Bhattacharya, B., Golin, M.J., Higashikawa, Y., Kameda, T., Katoh, N.: Improved algorithms for computing k-sink on dynamic flow path networks. In: Workshop on Algorithms and Data Structures. pp. 133--144. Springer (2017)

\bibitem{treeminmax}
Chen, D., Golin, M.: Minmax centered k-partitioning of trees and applications to sink evacuation with dynamic confluent flows. Algorithmica (in press)  (2022)

\bibitem{epstein2018min}
Epstein, L., Johnson, D.S., Levin, A.: Min-sum bin packing. Journal of Combinatorial Optimization  \textbf{36}(2),  508--531 (2018)

\bibitem{epstein2007minimum}
Epstein, L., Levin, A.: Minimum weighted sum bin packing. In: International Workshop on Approximation and Online Algorithms. pp. 218--231. Springer (2007)

\bibitem{ford1958constructing}
Ford~Jr, L.R., Fulkerson, D.R.: Constructing maximal dynamic flows from static flows. Operations research  \textbf{6}(3),  419--433 (1958)

\bibitem{fu2016clustering}
Fu, N., Suppakitpaisarn, V.: Clustering 1-dimensional periodic network using betweenness centrality. Computational Social Networks  \textbf{3}(1),  1--20 (2016)

\bibitem{fujie2021minmax}
Fujie, T., Higashikawa, Y., Katoh, N., Teruyama, J., Tokuni, Y.: Minmax regret 1-sink location problems on dynamic flow path networks with parametric weights. In: International Workshop on Algorithms and Computation. pp. 52--64. Springer (2021)

\bibitem{higashikawa2015minimax}
Higashikawa, Y., Augustine, J., Cheng, S.W., Golin, M.J., Katoh, N., Ni, G., Su, B., Xu, Y.: Minimax regret 1-sink location problem in dynamic path networks. Theoretical Computer Science  \textbf{588},  24--36 (2015)

\bibitem{higashikawa2014minimax}
Higashikawa, Y., Golin, M.J., Katoh, N.: Minimax regret sink location problem in dynamic tree networks with uniform capacity. In: International Workshop on Algorithms and Computation. pp. 125--137. Springer (2014)

\bibitem{higashikawa2019survey}
Higashikawa, Y., Katoh, N.: A survey on facility location problems in dynamic flow networks. The Review of Socionetwork Strategies  \textbf{13}(2),  163--208 (2019)

\bibitem{higashikawa2022almost}
Higashikawa, Y., Katoh, N., Teruyama, J.: Almost linear time algorithms for some problems on dynamic flow networks. In: Sublinear Computation Paradigm, pp. 65--85. Springer, Singapore (2022)

\bibitem{kamiyama2006efficient}
Kamiyama, N., Katoh, N., Takizawa, A.: An efficient algorithm for evacuation problem in dynamic network flows with uniform arc capacity. IEICE transactions on information and systems  \textbf{89}(8),  2372--2379 (2006)

\bibitem{karp1972reducibility}
Karp, R.M.: Reducibility among combinatorial problems. In: Complexity of computer computations, pp. 85--103. Springer (1972)

\bibitem{klinz2004minimum}
Klinz, B., Woeginger, G.J.: Minimum-cost dynamic flows: The series-parallel case. Networks: An International Journal  \textbf{43}(3),  153--162 (2004)

\bibitem{rhee1987probabilistic}
Rhee, W.T.: Probabilistic analysis of the next fit decreasing algorithm for bin-packing. Operations research letters  \textbf{6}(4),  189--191 (1987)

\end{thebibliography}

\section*{Appendix}

\subsection{NP-Hardness of MS-DWSF}
We show that MS-DWSF is NP-hard by a reduction to the partition problem in the following theorem.
\begin{theorem}
MS-DWSF is NP-hard even when the input path graph has two nodes.
\end{theorem}
\begin{proof}
Recall that, in the partition problem \cite{karp1972reducibility}, we have $\mathsf{m}$ items, denoted by $\{1, \dots, \mathsf{m}\}$. 
The size of items $i \in \{1, \dots, \mathsf{m}\}$ is $\mathsf{s}(i) \in \mathbb{Z}_{+}$. 
Suppose that $\sum\limits_i \mathsf{s}(i) = 2\mathsf{C}$. 
We aim to answer if there is $\mathsf{S} \subseteq \{1, \dots, \mathsf{m}\}$ such that $\sum\limits_{i \in \mathsf{S}} \mathsf{s}(i) = \mathsf{C}$. It is known that the partition problem is NP-hard.

Now, let us consider an instance of the MS-DWSF such that there are two nodes $\{1,2\}$ on the path graph, and the facility is located at node 2. The number of groups at node 1 (denoted by $m_1$) is $\mathsf{m}$. We also have $S(G_{1,i}) = w(G_{1,i}) = \mathsf{s}(i)$ for all $1 \leq i \leq m_1$, $C = \mathsf{C}$, and $d(\{1,2\}) = 1$. 

Since $\sum\limits_i S(G_{1,i}) = 2\mathsf{C} = 2C$, if there is $\mathsf{S}$ such that $\sum\limits_{i\in S} S(G_{1,i}) = \mathsf{C} = C$, we can send all the groups in two units of time by setting $D_{1}^{(1)} = \{G_{1,i} : i \in \mathsf{S}\}$ and $D_{1}^{(2)} = \{G_{i,1}, \dots, G_{i,m_i}\} \backslash D_{1}^{(1)}$. It is clear that those $D_1^{(1)}, D_1^{(2)}$ are the optimal solution as any other sets would give larger objective values. 

If there exists $\mathsf{S}$ such that $\sum\limits_{i\in S} S(G_{1,i}) = \mathsf{C} = C$, the optimal value of MS-DWSF would be $3C$. If there is no such $S$, we cannot send all the groups in two unit times. The optimal value must be larger than $3C$. Hence, if we can solve the MS-DWSF problem, we can give an answer to the partition problem. \qed
\end{proof}

\end{document}